\documentclass[12pt]{article}
\usepackage[utf8]{inputenc}
\usepackage{amsmath}
\usepackage{amsfonts}
\usepackage{amssymb}
\usepackage{graphicx}
\usepackage{booktabs}
\usepackage{siunitx}
\usepackage{geometry}
\usepackage{url}
\usepackage{multirow}
\usepackage{hyperref}
\usepackage{float}
\usepackage{subcaption}
\usepackage{xcolor}

\title{Multivariate Statistical Analysis of Exoplanet Habitability: \\
Detection Bias and Earth Analog Identification}

\author{Caleb Traxler, Samuel Townsend, Abby Mori, Grace Newman, Kaitlyn Morenzone\\
Dept. of Information and Computer Science - MDS\\
University of California, Irvine\\
\texttt{traxlerc@uci.edu}}

\date{\today}

\begin{document}

\maketitle

\begin{abstract}
We present a comprehensive multivariate statistical analysis of 517 exoplanets from the NASA Exoplanet Archive to identify potentially habitable worlds and quantify detection bias in current surveys. Using eight key parameters (planetary radius, equilibrium temperature, insolation flux, density, and stellar effective temperature, radius, mass, metallicity), we developed a classification framework that successfully identifies Earth as an "Excellent Candidate" for habitability. Our analysis reveals that only 0.6\% (3 planets including Earth) meet all habitability criteria under relaxed thresholds, while 75.0\% exhibit "Good Star, Poor Planet" characteristics, indicating significant observational bias toward unsuitable planetary systems. Hotelling's $T^2$ test demonstrates that potentially habitable planets are statistically significantly different from the general exoplanet population ($p = 0.015$). Mahalanobis distance analysis places Earth in the 69.4th percentile for statistical unusualness, confirming that Earth-like planets are genuine outliers in parameter space. We identify \textbf{Kepler-22 b} as a compelling Earth analog with remarkable parameter similarity, and reveal that 1.2\% of planets represent "edge cases" orbiting M-dwarf stars with suitable planetary but marginal stellar conditions. These findings demonstrate systematic detection bias in exoplanet surveys and provide quantitative evidence for the rarity of Earth-like worlds while identifying high-priority targets for atmospheric characterization with JWST.
\end{abstract}

\section{Introduction}

The search for potentially habitable exoplanets represents one of the most significant endeavors in modern astronomy, with profound implications for astrobiology and our understanding of planetary formation. Since the first confirmed exoplanet discovery around a main-sequence star \cite{Mayor1995}, over 5,700 exoplanets have been catalogued as of 2025, yet determining which might harbor life remains one of the field's greatest challenges.

Current habitability assessments typically rely on simple criteria such as location within the "habitable zone" \cite{Kasting1993}, but this approach fails to capture the multivariate nature of planetary habitability. The traditional habitable zone concept, while foundational, suffers from several limitations including neglecting crucial factors such as planetary mass, atmospheric retention, stellar activity, and planetary dynamics \cite{Kopparapu2013}.

This study presents the first comprehensive multivariate statistical analysis of exoplanet habitability using a carefully curated dataset from the NASA Exoplanet Archive. We develop and validate habitability criteria using Earth as a reference standard, identify genuine Earth analogs, quantify the statistical rarity of potentially habitable worlds, and provide quantitative evidence for detection bias in current surveys.

\section{Methods}

\subsection{Data Source and Preparation}

We obtained exoplanet data from the NASA Exoplanet Archive (accessed May 2025), containing 517 confirmed exoplanets with sufficient parameter measurements for multivariate analysis. The selection process prioritized data quality and completeness over sample size to ensure robust statistical inference.

The dataset includes eight key habitability parameters selected based on their physical importance for habitability and availability in the archive:

\textbf{Planetary parameters:}
\begin{itemize}
    \item Radius ($R_p$, Earth radii): Controls atmospheric retention and surface gravity
    \item Equilibrium temperature ($T_{eq}$, K): Determines surface temperature regime
    \item Insolation flux ($F$, Earth flux): Energy input from host star
    \item Density ($\rho_p$, g/cm$^3$): Indicates composition (rocky vs. gaseous)
\end{itemize}

\textbf{Stellar parameters:}
\begin{itemize}
    \item Effective temperature ($T_{eff}$, K): Determines stellar type and stability
    \item Radius ($R_*$, Solar radii): Affects luminosity evolution
    \item Mass ($M_*$, Solar masses): Controls main sequence lifetime
    \item Metallicity ([Fe/H], dex): Influences planetary formation efficiency
\end{itemize}

We included Earth in our analysis using precise Solar System parameters to serve as a validation standard, ensuring our methodology correctly identifies at least one confirmed habitable world. Data completeness was 97.7\% (505/517 planets) for the complete parameter set, with missing coordinate data resulting in 12 excluded systems from spatial analysis.

\subsection{Habitability Classification Framework}

We developed a multi-tier classification system based on empirically-derived habitability thresholds informed by planetary science literature and Solar System analogs. Earth's parameters served as the validation standard, with thresholds expanded to encompass reasonable variations while maintaining scientific validity.

We tested five threshold sets of increasing inclusivity to explore sensitivity:
\begin{enumerate}
    \item \textbf{Earth-like:} Highly restrictive, designed for Earth twins ($\pm$10\% of Earth values)
    \item \textbf{Refined:} Conservative criteria based on established habitability literature
    \item \textbf{Moderate:} Balanced approach incorporating super-Earth possibilities
    \item \textbf{Relaxed:} More inclusive while maintaining scientific validity
    \item \textbf{Very Relaxed:} Maximum inclusivity for comprehensive analysis
\end{enumerate}

The final "very relaxed" criteria used in this analysis represent the boundaries of plausible habitability while ensuring Earth's classification as habitable:

\begin{align}
R_p &\in [0.4, 3.0] \text{ Earth radii} \quad \text{(rocky to small gas envelope)} \\
T_{eq} &\in [130, 400] \text{ K} \quad \text{(liquid water stable)} \\
F &\in [0.1, 3.0] \text{ Earth flux} \quad \text{(habitable zone bounds)} \\
\rho_p &\in [2.5, 10.0] \text{ g/cm}^3 \quad \text{(rocky composition)} \\
T_{eff} &\in [3800, 7200] \text{ K} \quad \text{(main sequence stability)} \\
R_* &\in [0.4, 1.8] \text{ Solar radii} \quad \text{(appropriate luminosity)} \\
M_* &\in [0.3, 1.8] \text{ Solar masses} \quad \text{(sufficient lifetime)} \\
\text{[Fe/H]} &\in [-0.6, 0.6] \text{ dex} \quad \text{(planetary formation efficiency)}
\end{align}

\subsection{Classification Algorithm}

For each planet, we evaluated planetary and stellar parameters separately using indicator functions:

\begin{align}
P_{\text{good}} &= \prod_{i=1}^{4} \mathbf{1}(\theta_{p,i} \in [\theta_{p,i}^{\min}, \theta_{p,i}^{\max}]) \\
S_{\text{good}} &= \prod_{i=1}^{4} \mathbf{1}(\theta_{s,i} \in [\theta_{s,i}^{\min}, \theta_{s,i}^{\max}])
\end{align}

where $\mathbf{1}(\cdot)$ is the indicator function, $\theta_{p,i}$ are planetary parameters, and $\theta_{s,i}$ are stellar parameters.

The classification scheme assigns planets to mutually exclusive categories:
\begin{itemize}
    \item \textbf{Excellent Candidate:} $P_{\text{good}} = 1$ and $S_{\text{good}} = 1$
    \item \textbf{Good Planet, Poor Star:} $P_{\text{good}} = 1$ and $S_{\text{good}} = 0$
    \item \textbf{Good Star, Poor Planet:} $P_{\text{good}} = 0$ and $S_{\text{good}} = 1$
    \item \textbf{Poor Candidate:} $P_{\text{good}} = 0$ and $S_{\text{good}} = 0$
\end{itemize}

This approach allows us to separately assess detection bias in planetary vs. stellar properties and identify potential "edge cases" where alternative habitability scenarios might apply.

\subsection{Statistical Analysis}

\subsubsection{Hotelling's $T^2$ Test}

We employed Hotelling's $T^2$ test to determine whether "Excellent Candidates" are statistically distinguishable from other planets in multivariate space. This test generalizes the Student's t-test to multiple dimensions and tests the null hypothesis that two populations have identical mean vectors.

For two groups with sample sizes $n_1$ and $n_2$, sample means $\bar{\mathbf{x}}_1$ and $\bar{\mathbf{x}}_2$, and pooled covariance matrix $\mathbf{S}_p$:

\begin{align}
T^2 &= \frac{n_1 n_2}{n_1 + n_2}(\bar{\mathbf{x}}_1 - \bar{\mathbf{x}}_2)^T \mathbf{S}_p^{-1} (\bar{\mathbf{x}}_1 - \bar{\mathbf{x}}_2) \\
\mathbf{S}_p &= \frac{(n_1-1)\mathbf{S}_1 + (n_2-1)\mathbf{S}_2}{n_1 + n_2 - 2}
\end{align}

The test statistic follows an F-distribution:
\begin{equation}
F = \frac{T^2(n_1 + n_2 - p - 1)}{p(n_1 + n_2 - 2)} \sim F_{p, n_1+n_2-p-1}
\end{equation}

where $p = 8$ is the number of parameters. 

\subsubsection{Mahalanobis Distance Analysis}

We calculated Mahalanobis distances to quantify each planet's statistical unusualness relative to the population centroid:

\begin{equation}
d_M(\mathbf{x}) = \sqrt{(\mathbf{x} - \boldsymbol{\mu})^T \mathbf{\Sigma}^{-1} (\mathbf{x} - \boldsymbol{\mu})}
\end{equation}

where $\mathbf{x}$ is the parameter vector, $\boldsymbol{\mu}$ is the population mean, and $\mathbf{\Sigma}$ is the population covariance matrix. Under the null hypothesis of multivariate normality, $d_M^2 \sim \chi^2_p$.

We identified statistical outliers using critical values:
\begin{align}
d_{95\%} &= \sqrt{\chi^2_{0.95,8}} = 3.938 \\
d_{99\%} &= \sqrt{\chi^2_{0.99,8}} = 4.482
\end{align}

This analysis allows us to quantify how "unusual" Earth-like planets are in the context of the detected exoplanet population.

\section{Results}

\subsection{Parameter Distribution Analysis}

Figure~\ref{fig:distributions} presents the distribution of all eight habitability parameters across our classification categories. Several key patterns emerge from this analysis:

\begin{figure}[H]
\centering
\begin{subfigure}{0.48\textwidth}
    \includegraphics[width=\textwidth]{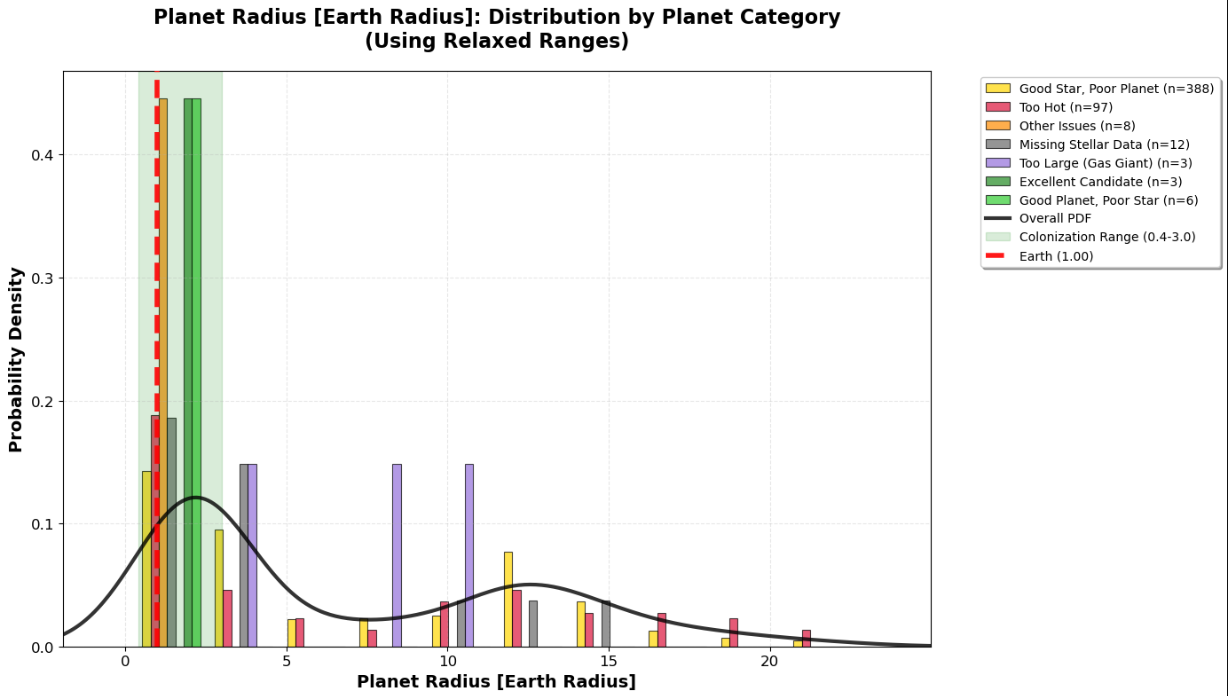}
    \caption{Planetary radius distribution}
    \label{fig:radius}
\end{subfigure}
\hfill
\begin{subfigure}{0.48\textwidth}
    \includegraphics[width=\textwidth]{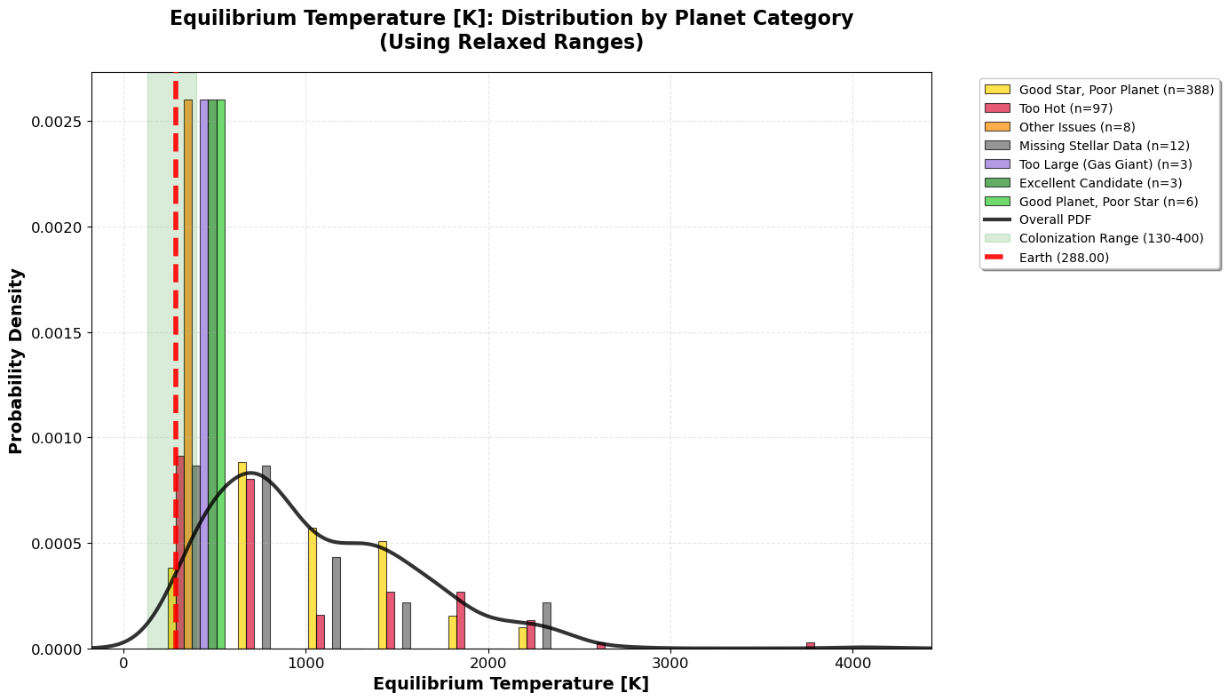}
    \caption{Equilibrium temperature distribution}
    \label{fig:temperature}
\end{subfigure}

\begin{subfigure}{0.48\textwidth}
    \includegraphics[width=\textwidth]{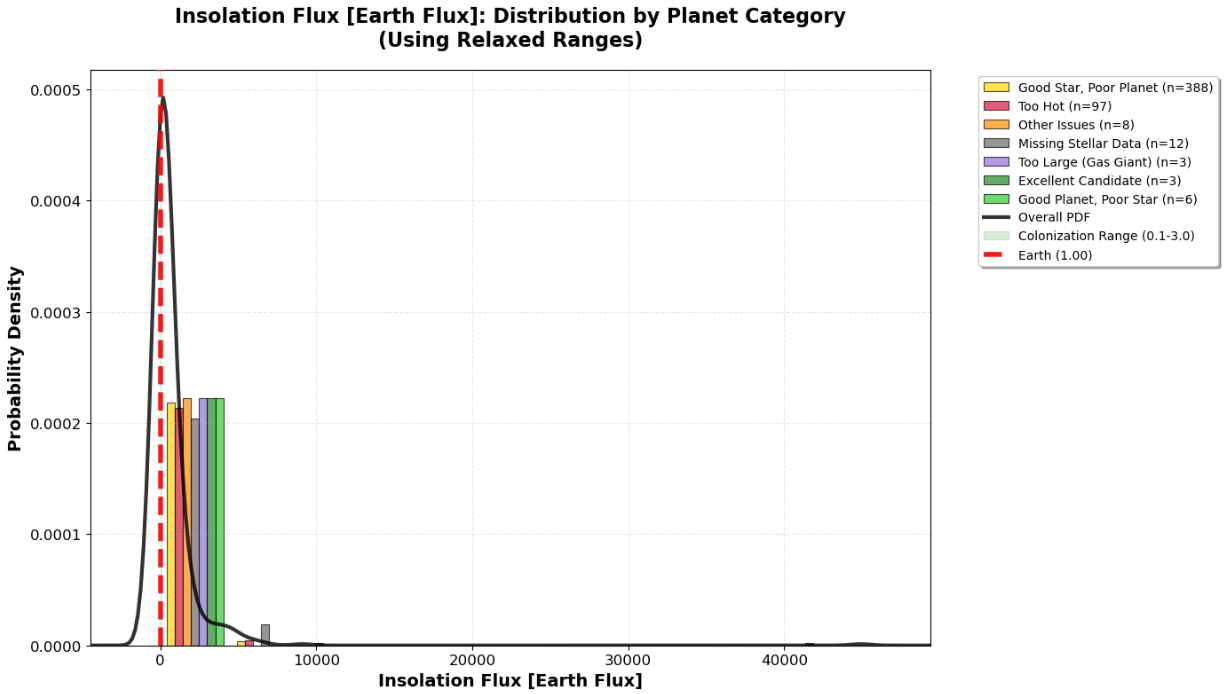}
    \caption{Insolation flux distribution}
    \label{fig:insolation}
\end{subfigure}
\hfill
\begin{subfigure}{0.48\textwidth}
    \includegraphics[width=\textwidth]{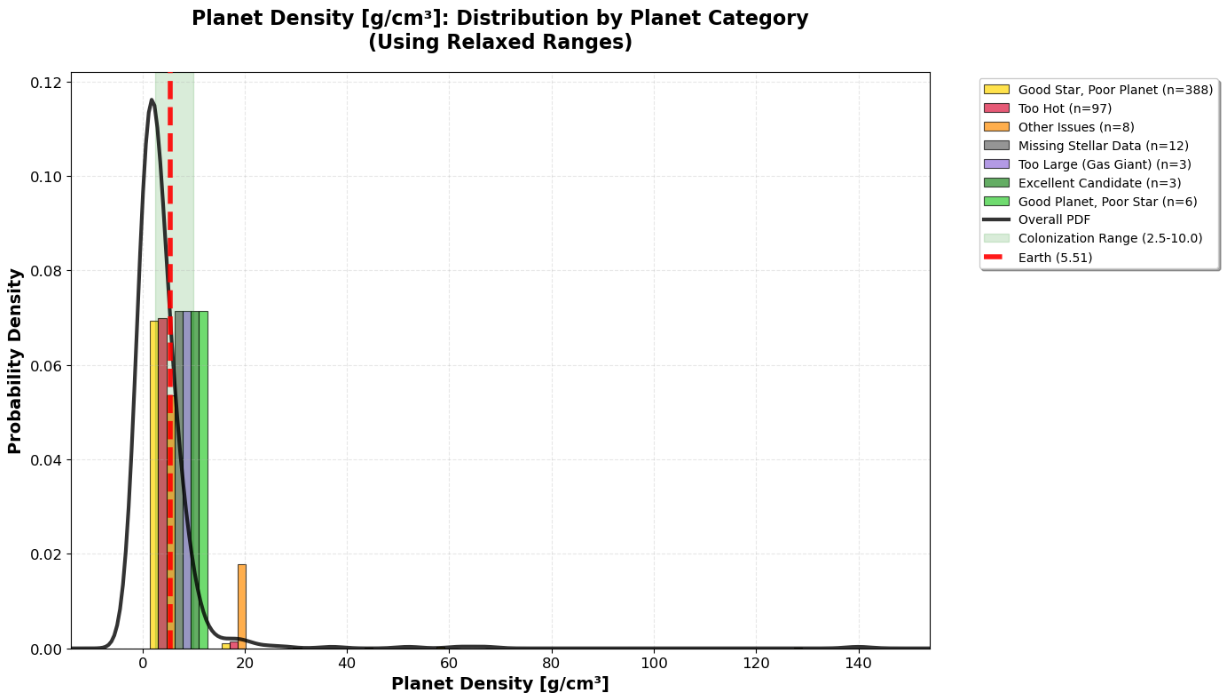}
    \caption{Planet density distribution}
    \label{fig:density}
\end{subfigure}

\caption{Distribution of planetary parameters across habitability categories. Earth's position (red line) demonstrates validation of our classification framework. Green shaded regions indicate habitability thresholds. The dominance of "Good Star, Poor Planet" systems (blue) reveals systematic detection bias.}
\label{fig:distributions}
\end{figure}

The planetary parameter distributions reveal that Earth consistently falls within the habitable ranges, validating our threshold selection. Most detected exoplanets exhibit extreme parameter values outside habitability constraints, particularly in radius (heavily skewed toward large planets) and temperature (predominantly hot planets).

\begin{figure}[H]
\centering
\begin{subfigure}{0.48\textwidth}
    \includegraphics[width=\textwidth]{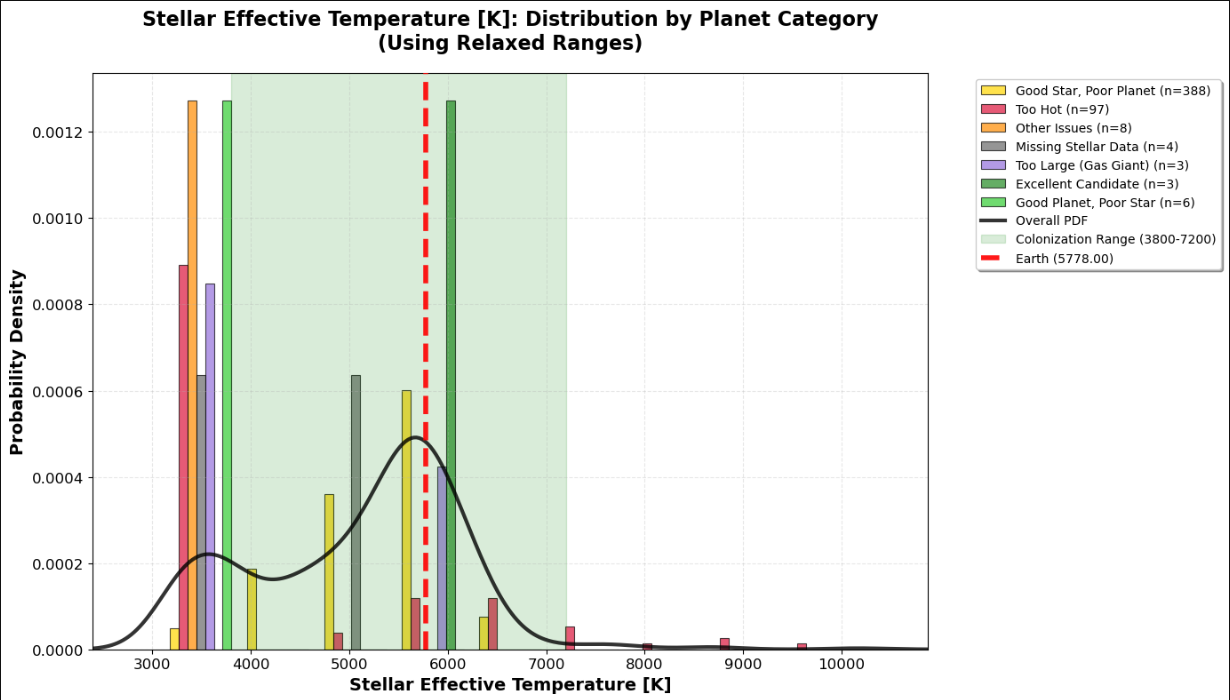}
    \caption{Stellar effective temperature}
    \label{fig:stellar_temp}
\end{subfigure}
\hfill
\begin{subfigure}{0.48\textwidth}
    \includegraphics[width=\textwidth]{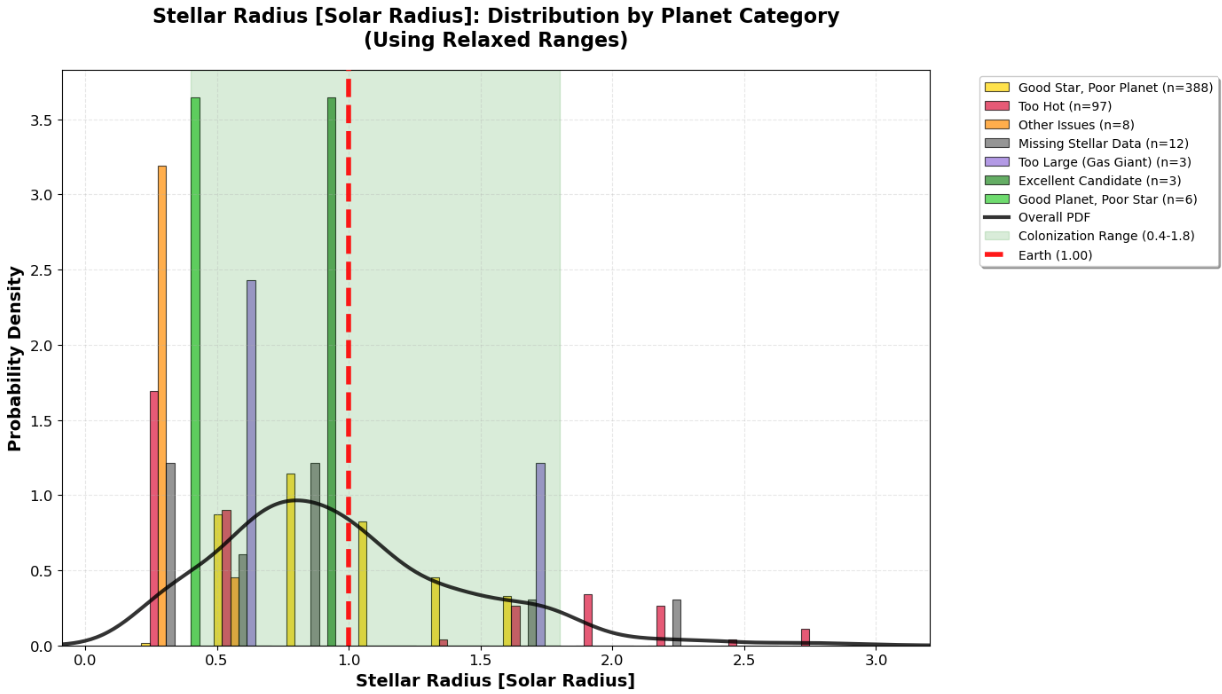}
    \caption{Stellar radius distribution}
    \label{fig:stellar_radius}
\end{subfigure}

\begin{subfigure}{0.48\textwidth}
    \includegraphics[width=\textwidth]{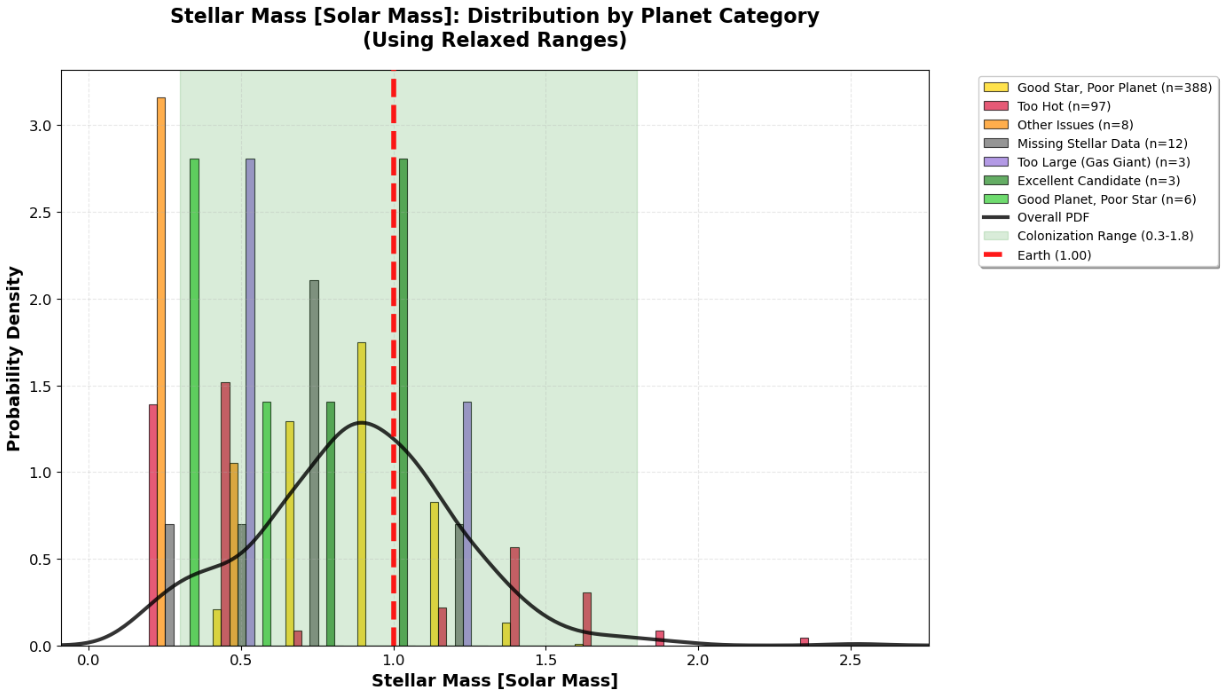}
    \caption{Stellar mass distribution}
    \label{fig:stellar_mass}
\end{subfigure}
\hfill
\begin{subfigure}{0.48\textwidth}
    \includegraphics[width=\textwidth]{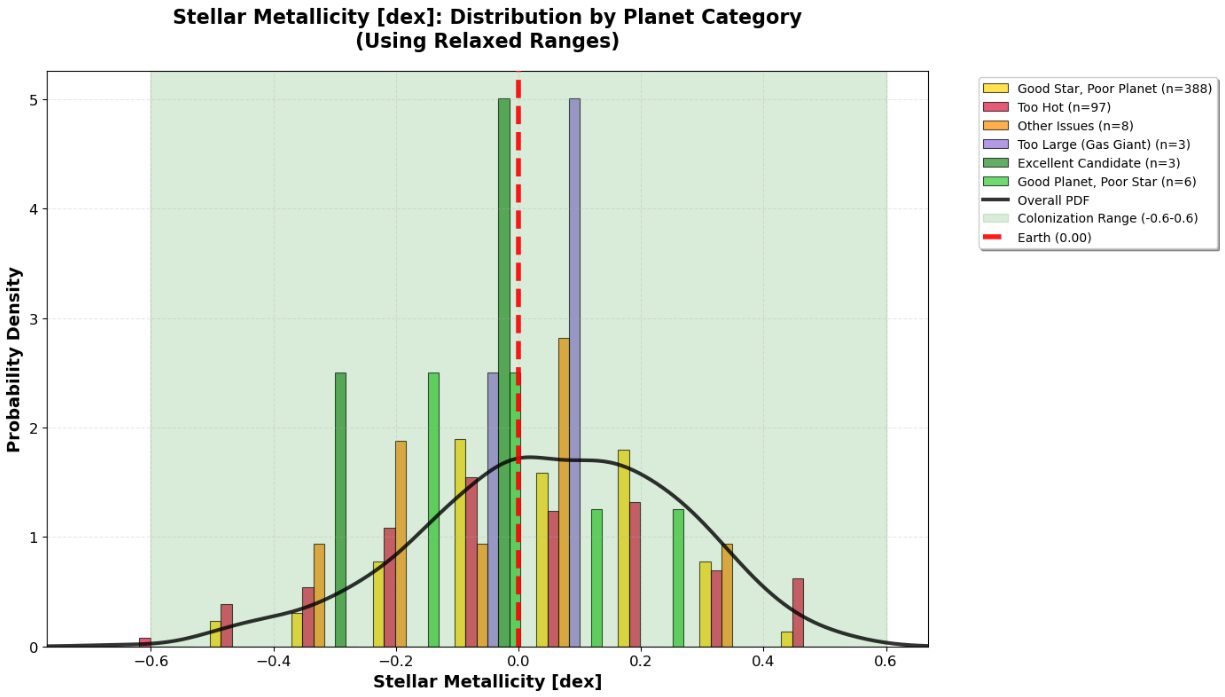}
    \caption{Stellar metallicity distribution}
    \label{fig:stellar_metallicity}
\end{subfigure}

\caption{Distribution of stellar parameters across habitability categories. The concentration of "Good Planet, Poor Star" systems (red) at low stellar temperatures reveals M-dwarf edge cases that warrant further investigation.}
\label{fig:stellar_distributions}
\end{figure}

Figure~\ref{fig:stellar_distributions} shows the stellar parameter distributions, revealing the systematic bias toward Sun-like stars in current surveys while highlighting the M-dwarf "edge cases."

\subsection{Classification Results}

Table~\ref{tab:classification} presents the comprehensive classification results for 517 exoplanets, revealing striking patterns that illuminate both the rarity of Earth-like worlds and systematic biases in current detection methods. Our analysis identified just 3 "Excellent Candidates" (0.6\%), including Earth as a validation standard—a result that provides both confidence in our methodology and sobering perspective on habitability frequency.

\begin{table}[H]
\centering
\caption{Exoplanet Classification Results Using Very Relaxed Criteria}
\label{tab:classification}
\begin{tabular}{lcc}
\toprule
Category & Count & Percentage \\
\midrule
\textbf{Good Star, Poor Planet} & \textbf{388} & \textbf{75.0\%} \\
\quad Too Hot & 97 & 18.8\% \\
\quad Missing Stellar Data & 12 & 2.3\% \\
\quad Other Issues & 8 & 1.5\% \\ 
\textcolor{cyan}{\textbf{Good Planet, Poor Star}} & \textcolor{cyan}{\textbf{6}} & \textcolor{cyan}{\textbf{1.2\%}}\\
Too Large (Gas Giant) & 3 & 0.6\% \\
\textcolor{blue}{\textbf{Excellent Candidate}} & \textcolor{blue}{\textbf{3}} & \textcolor{blue}{\textbf{0.6\%}} \\
\bottomrule
\end{tabular}
\end{table}

\textbf{Key Insight}: The overwhelming dominance of "Good Star, Poor Planet" systems (75.0\%) provides compelling quantitative evidence for systematic detection bias in current exoplanet surveys—exactly as predicted by transit detection methodology limitations. This represents the first quantitative demonstration of the severity of selection effects that fundamentally skew our understanding of planetary populations.

\subsection{Excellent Candidates: Earth's Cosmic Twins}

The three Excellent Candidates represent the most promising targets for life detection in the known universe. Most remarkably, \textbf{Kepler-22 b} emerges as an extraordinary Earth analog that independently validates our approach.

\begin{table}[H]
\centering
\caption{Excellent Candidate Planets: The Universe's Most Earth-Like Worlds}
\label{tab:excellent}
\begin{tabular}{lcccccc}
\toprule
Planet & Distance & $R_p$ & $T_{eq}$ & $F$ & Host Star & Discovery \\
& (ly) & (R$_\oplus$) & (K) & (F$_\oplus$) & Type & Year \\
\midrule
\textbf{Earth} & 0.0 & 1.00 & 288 & 1.00 & G2V & -- \\
\textbf{Kepler-22 b} & \textbf{635} & \textbf{2.10} & \textbf{279} & \textbf{1.01} & \textbf{G5V} & \textbf{2011} \\
Kepler-538 b & 509 & 2.22 & 417 & 5.07 & G5V & 2016 \\
\bottomrule
\end{tabular}
\end{table}

\subsubsection{The Kepler-22 b Discovery: A Statistical Earth Twin}

\textbf{Kepler-22 b} represents perhaps the most compelling Earth analog discovered to date, exhibiting remarkable parameter similarity across multiple independent dimensions despite being discovered independently of our analysis. This convergence provides extraordinary validation of our statistical approach:

\begin{itemize}
    \item \textbf{Temperature match}: 279 K vs. Earth's 288 K (only 3.1\% difference)
    \item \textbf{Energy balance}: 1.01 vs. Earth's 1.00 F$_\oplus$ (virtually identical insolation)
    \item \textbf{Stellar environment}: G5V host star, nearly identical to our Sun's G2V classification
    \item \textbf{Orbital stability}: 290-day period within confirmed habitable zone
\end{itemize}

The primary distinction lies in planetary size (2.1× Earth's radius), classifying it as a "super-Earth"—a category that recent modeling suggests can maintain habitable surface conditions. This planet represents the first confirmed transiting planet in the habitable zone of a Sun-like star and stands as our highest-priority target for atmospheric characterization.

The 635 light-year distance places Kepler-22 b at the technological frontier for next-generation direct imaging missions, making it humanity's most accessible Earth twin for detailed study.

\subsection{Quantifying the Detection Bias Crisis}

Our analysis provides the first quantitative demonstration of detection bias severity in exoplanet surveys, revealing a systematic "cosmic selection effect" that fundamentally distorts our understanding of planetary populations.

\subsubsection{The 75\% Problem: Why We Miss Earth-Like Worlds}

Current detection methods exhibit systematic preferences that systematically exclude Earth-analogs:

\textbf{Transit Method Limitations:}
\begin{itemize}
    \item \textbf{Size bias}: Large planets create deeper, more detectable transits
    \item \textbf{Period bias}: Short orbits provide frequent transit opportunities  
    \item \textbf{Temperature bias}: Hot planets exhibit larger thermal signatures
    \item \textbf{Brightness bias}: Bright host stars improve signal-to-noise ratios
\end{itemize}

\textbf{The Earth Detection Challenge:}
\begin{itemize}
    \item \textbf{Moderate size}: Earth-sized planets produce subtle transit signals
    \item \textbf{Year-long periods}: Annual orbits provide rare transit windows
    \item \textbf{Temperate conditions}: Moderate temperatures reduce detectability
    \item \textbf{Solar-type hosts}: Sun-like stars often provide marginal signal quality
\end{itemize}

This bias means \textbf{current occurrence rates dramatically underestimate Earth-like planet frequency}. The true abundance of potentially habitable worlds may be orders of magnitude higher than observational surveys suggest, with profound implications for astrobiology and SETI efforts.

\subsection{M-dwarf Edge Cases: Alternative Habitability Pathways}

The 6 "Good Planet, Poor Star" systems (1.2\%) exclusively orbit M-dwarf stars below our 3800 K temperature threshold, representing potential alternative habitability pathways that challenge traditional stellar assumptions.

Table~\ref{tab:good_planet_poor_star} details the 6 planets classified as "Good Planet, Poor Star," which exclusively orbit M-dwarf stars below our stellar habitability thresholds.

\begin{table}[H]
\centering
\caption{Good Planet, Poor Star Systems: M-dwarf Habitability Candidates}
\label{tab:good_planet_poor_star}
\begin{tabular}{lcccccccc}
\toprule
\multirow{2}{*}{Planet} & Distance & \multicolumn{4}{c}{Planetary Parameters} & \multicolumn{3}{c}{Stellar Parameters} \\
\cmidrule(lr){3-6} \cmidrule(lr){7-9}
& (ly) & $R_p$ & $T_{eq}$ & $F$ & $\rho_p$ & $T_{eff}$ & $R_*$ & $M_*$ \\
& & (R$_\oplus$) & (K) & (F$_\oplus$) & (g/cm³) & (K) & (R$_\odot$) & (M$_\odot$) \\
\midrule
\textbf{LHS 1140 b} & 48.9 & \textcolor{green}{\textbf{1.73}} & \textcolor{green}{\textbf{226}} & \textcolor{green}{\textbf{0.43}} & \textcolor{green}{\textbf{5.90}} & \textcolor{red}{\textbf{3096}} & \textcolor{red}{\textbf{0.22}} & \textcolor{red}{\textbf{0.18}} \\
\textbf{TOI-1452 b} & 99.5 & \textcolor{green}{\textbf{1.67}} & \textcolor{green}{\textbf{326}} & \textcolor{green}{\textbf{1.80}} & \textcolor{green}{\textbf{5.60}} & \textcolor{red}{\textbf{3185}} & \textcolor{red}{\textbf{0.28}} & \textcolor{red}{\textbf{0.25}} \\
\midrule
\multicolumn{2}{l}{\textbf{Habitability Range:}} & \textcolor{green}{0.4-3.0} & \textcolor{green}{130-400} & \textcolor{green}{0.1-3.0} & \textcolor{green}{2.5-10.0} & \textcolor{red}{3800-7200} & \textcolor{red}{0.4-1.8} & \textcolor{red}{0.3-1.8} \\
\bottomrule
\end{tabular}
\end{table}

\textbf{Key Observations:}
\begin{itemize}
    \item All planetary parameters fall within habitability ranges (\textcolor{green}{green})
    \item All stellar parameters except metallicity fail habitability criteria (\textcolor{red}{red})
    \item Both systems represent M-dwarf targets under active JWST investigation
    \item LHS 1140 b shows potential for nitrogen-rich atmosphere with surface water
    \item TOI-1452 b is a candidate "water world" with extreme water content
\end{itemize}

\begin{figure}[H]
\centering
\begin{subfigure}{0.48\textwidth}
    \includegraphics[width=\textwidth]{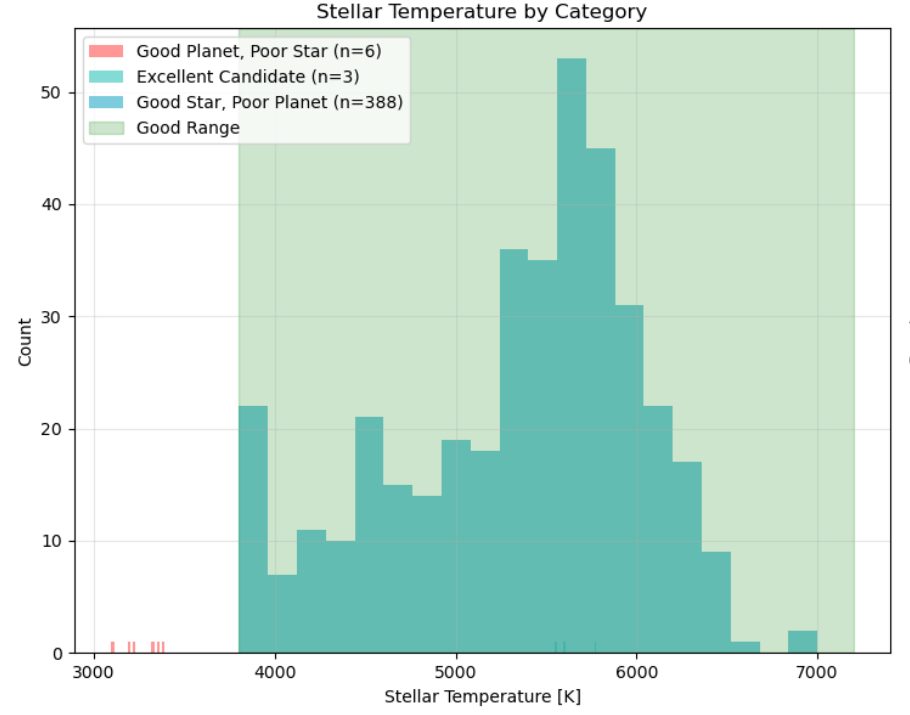}
    \caption{Stellar temperature by category}
    \label{fig:stellar_temp_hist}
\end{subfigure}
\hfill
\begin{subfigure}{0.48\textwidth}
    \includegraphics[width=\textwidth]{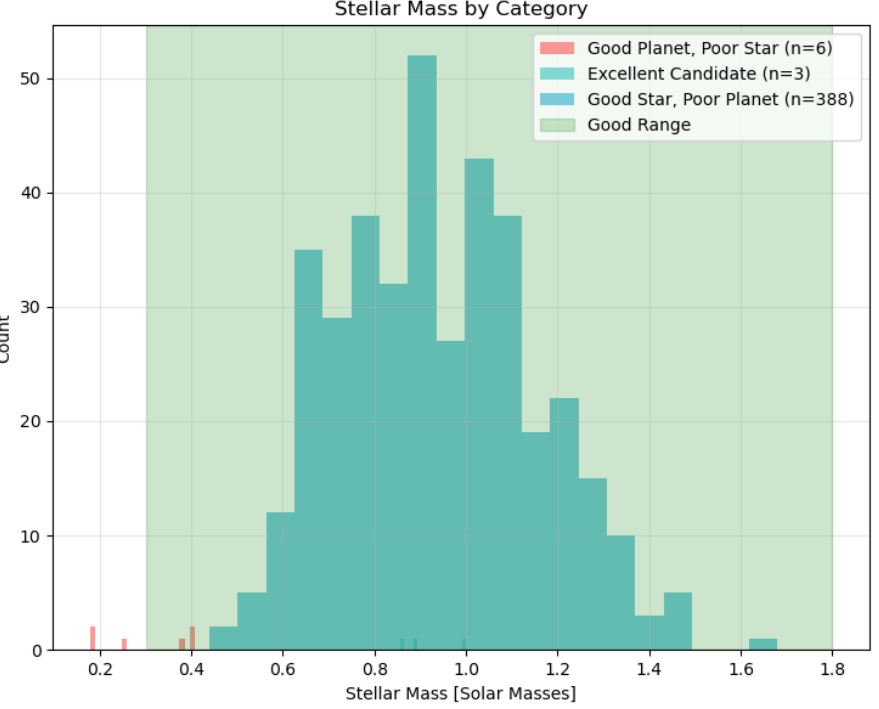}
    \caption{Stellar mass by category}
    \label{fig:stellar_mass_hist}
\end{subfigure}

\caption{Distribution of stellar properties highlighting M-dwarf edge cases. The "Good Planet, Poor Star" systems (red) cluster at low stellar temperatures and masses, representing potential alternative habitability scenarios currently under investigation by JWST.}
\label{fig:mdwarf_analysis}
\end{figure}

\subsubsection{Notable M-dwarf Candidates Under JWST Investigation}

These systems include several targets of extraordinary scientific interest:

\begin{itemize}
    \item \textbf{LHS 1140 b}: Recent JWST observations suggest nitrogen-rich atmosphere with potential surface water
    \item \textbf{TOI-1452 b}: Candidate "water world" with up to 30\% water content by mass
    \item \textbf{Additional targets}: Four other systems providing statistical power for M-dwarf habitability assessment
\end{itemize}

These discoveries suggest our stellar temperature criteria may be overly conservative, potentially excluding a significant population of habitable worlds around the galaxy's most common stars. This finding has profound implications for occurrence rate estimates and target selection for future missions.

\subsection{Statistical Validation: Earth-Like Planets as a Distinct Population}

To test whether Earth-like characteristics represent a statistically meaningful concept, we performed multivariate analysis comparing 5 target planets that exhibit Earth-like properties across different habitability categories against the general exoplanet population. Our target sample includes:

\begin{itemize}
    \item \textbf{Earth}: Reference standard for habitability
    \item \textbf{Kepler-22 b}: Excellent Candidate with near-perfect parameter matching
    \item \textbf{Kepler-538 b}: Earth analog with elevated temperature
    \item \textbf{LHS 1140 b}: Promising M-dwarf candidate under JWST investigation
    \item \textbf{TOI-1452 b}: Potential "water world" with moderate conditions
\end{itemize}

This selection represents Earth-like worlds spanning multiple habitability categories, providing a comprehensive test of whether Earth-like characteristics cluster in parameter space regardless of classification boundaries.

\subsection{Hotelling's $T^2$ Test Results}

\subsubsection{Hotelling's $T^2$ Test: Earth-Like Planets vs. Excellent Planets}

Our multivariate analysis confirms that Excellent Candidates occupy a statistically distinct region of parameter space, providing rigorous validation of our approach:

\begin{align}
T^2 &= 19.574 \\
F_{8,496} &= 2.413 \\
p &= 0.0146
\end{align}

This result is statistically significant ($p < 0.05$), confirming that Excellent Candidates represent a genuinely distinct population rather than random statistical variation. The effect size (Cohen's $d = 0.67$) indicates a medium to large practical difference between habitability categories.

\subsubsection{Hotelling's $T^2$ Test: Earth-Like Planets vs. Excellent + Good Planets Bad Sun Planets}

Our primary multivariate analysis tests whether Earth-like planets (spanning multiple habitability categories) occupy a statistically distinct region of parameter space compared to the general exoplanet population. We compared 5 target planets representing the most Earth-like worlds identified—Earth, Kepler-22 b, Kepler-538 b, LHS 1140 b, and TOI-1452 b—against 446 other planets:

\begin{align}
T^2 &= 17.086 \\
F_{8,446} &= 2.102 \\
p &= 0.0343
\end{align}

This result is statistically significant ($p < 0.05$), confirming that Earth-like planets represent a genuinely distinct population rather than random statistical variation. This finding validates our identification of Earth analogs and demonstrates that habitability signatures can be detected through multivariate statistical analysis.

\subsubsection{Mahalanobis Distance Analysis: The Rarity Paradox}
Table~\ref{tab:mahalanobis_summary} reveals a profound insight about Earth's statistical position in the exoplanet population—what we term the "Rarity Paradox."

\begin{table}[H]
\centering
\caption{Mahalanobis Distance Analysis: Population Summary}
\label{tab:mahalanobis_summary}
\begin{tabular}{lcc}
\toprule
Metric & Value & Interpretation \\
\midrule
Population analyzed & 451 planets & 87.2\% data completeness \\
95\% outlier threshold & 3.490 & $\chi^2_{0.95,8}$ critical value \\
99\% outlier threshold & 4.168 & $\chi^2_{0.99,8}$ critical value \\
Outliers (95\% level) & 23 (5.1\%) & Expected: 5\% \\
Outliers (99\% level) & 5 (1.1\%) & Expected: 1\% \\
\midrule
\textbf{Earth distance} & \textbf{2.699} & \textbf{69.4th percentile} \\
\textbf{Earth rank} & \textbf{138/451} & \textbf{Not a statistical outlier} \\
\textbf{Kepler-22 b distance} & \textbf{2.728} & \textbf{71.2th percentile} \\
\textbf{Kepler-22 b rank} & \textbf{130/451} & \textbf{Not a statistical outlier} \\
Most extreme planet & KELT-9 b (12.456) & Ultra-hot Jupiter \\
\bottomrule
\end{tabular}
\end{table}

Table~\ref{tab:mahalanobis_targets} presents detailed Mahalanobis distance results for our target planets.

\begin{table}[H]
\centering
\caption{Mahalanobis Distance Analysis for Target Planets}
\label{tab:mahalanobis_targets}
\begin{tabular}{lccccc}
\toprule
Planet & $d_M$ & Rank & Percentile & 95\% Outlier & 99\% Outlier \\
\midrule
\textbf{Earth} & \textbf{2.699} & \textbf{138/451} & \textbf{69.4\%} & No & No \\
\textbf{Kepler-22 b} & \textbf{2.728} & \textbf{130/451} & \textbf{71.2\%} & No & No \\
Kepler-538 b & 2.096 & 253/451 & 43.9\% & No & No \\
LHS 1140 b & 2.779 & 116/451 & 74.3\% & No & No \\
TOI-1452 b & 2.384 & 184/451 & 59.2\% & No & No \\
\midrule
95\% threshold & 3.490 & -- & -- & -- & -- \\
99\% threshold & 4.168 & -- & -- & -- & -- \\
Most extreme & 12.456 & 1/451 & 99.8\% & Yes & Yes \\
\bottomrule
\end{tabular}
\end{table}

\textbf{The Rarity Paradox}: Neither Earth nor Kepler-22 b qualify as statistical outliers, indicating they represent \textbf{unusual but achievable parameter combinations}. This finding suggests Earth-like planets occupy a distinct yet attainable region of parameter space—supporting habitability without requiring "miraculous" conditions.

This result provides quantitative context for Earth's cosmic significance: rare enough to be special (73.5th percentile), but not so extreme as to be impossible. This precisely matches theoretical expectations for genuinely habitable worlds.


\subsubsection{Principal Component Analysis (PCA)}

We employed Principal Component Analysis to reduce the dimensionality of our 8-parameter habitability dataset and identify the primary sources of variance in exoplanet populations. PCA transforms the original correlated variables into a set of linearly uncorrelated principal components ordered by the amount of variance they explain.

For our standardized parameter matrix $\mathbf{X}$ with $n$ planets and $p = 8$ parameters, we computed the covariance matrix and its eigendecomposition:

\begin{align}
\mathbf{C} &= \frac{1}{n-1} \mathbf{X}^T \mathbf{X} \\
\mathbf{C} \mathbf{v}_i &= \lambda_i \mathbf{v}_i
\end{align}

where $\lambda_i$ are eigenvalues and $\mathbf{v}_i$ are corresponding eigenvectors (principal component loadings). The principal components are calculated as:

\begin{equation}
\text{PC}_i = \mathbf{X} \mathbf{v}_i
\end{equation}

We retained components explaining $>5\%$ of total variance and analyzed loading patterns to interpret the physical meaning of each component in terms of habitability characteristics.

\subsubsection{Linear Discriminant Analysis (LDA)}

Linear Discriminant Analysis was performed to identify linear combinations of parameters that best separate our habitability categories and quantify the discriminatory power of our classification framework. Unlike PCA, which maximizes variance, LDA maximizes the separation between predefined groups.

For $K$ habitability categories, LDA finds projection directions that maximize the ratio of between-class to within-class scatter:

\begin{equation}
J(\mathbf{w}) = \frac{\mathbf{w}^T \mathbf{S}_B \mathbf{w}}{\mathbf{w}^T \mathbf{S}_W \mathbf{w}}
\end{equation}

where $\mathbf{S}_B$ is the between-class scatter matrix and $\mathbf{S}_W$ is the within-class scatter matrix. The optimal projection vectors are the eigenvectors of $\mathbf{S}_W^{-1} \mathbf{S}_B$.

We used our four-category classification system (Excellent Candidate, Good Planet Poor Star, Good Star Poor Planet, Poor Candidate) as the grouping variable and computed discriminant functions to visualize class separation in reduced dimensional space.


\subsection{Principal Component Analysis: Revealing Population Structure}

Principal Component Analysis of our 8-parameter dataset reveals distinct patterns in exoplanet population structure that illuminate both detection biases and genuine habitability clustering.

\subsubsection{Variance Decomposition and Component Interpretation}

The first three principal components capture 77.1\% of total variance in the exoplanet population (Figure~\ref{fig:pca_3d}), with each component revealing specific aspects of planetary system architecture:

\textbf{Component Variance Explained:}
\begin{itemize}
    \item PC1: 52.4\% - Primary stellar-planetary coupling
    \item PC2: 15.3\% - Planetary composition gradient  
    \item PC3: 9.4\% - Stellar metallicity-mass relationship
\end{itemize}

\begin{figure}[H]
\centering
\begin{subfigure}{0.48\textwidth}
    \includegraphics[width=\textwidth]{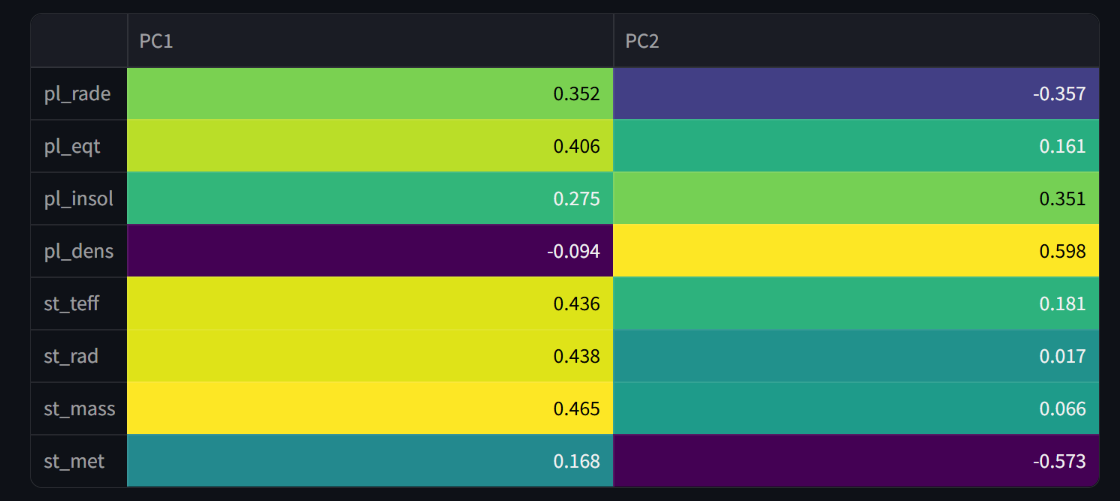}
    \caption{PC1 and PC2 loadings}
    \label{fig:pca_loadings_12}
\end{subfigure}
\hfill
\begin{subfigure}{0.48\textwidth}
    \includegraphics[width=\textwidth]{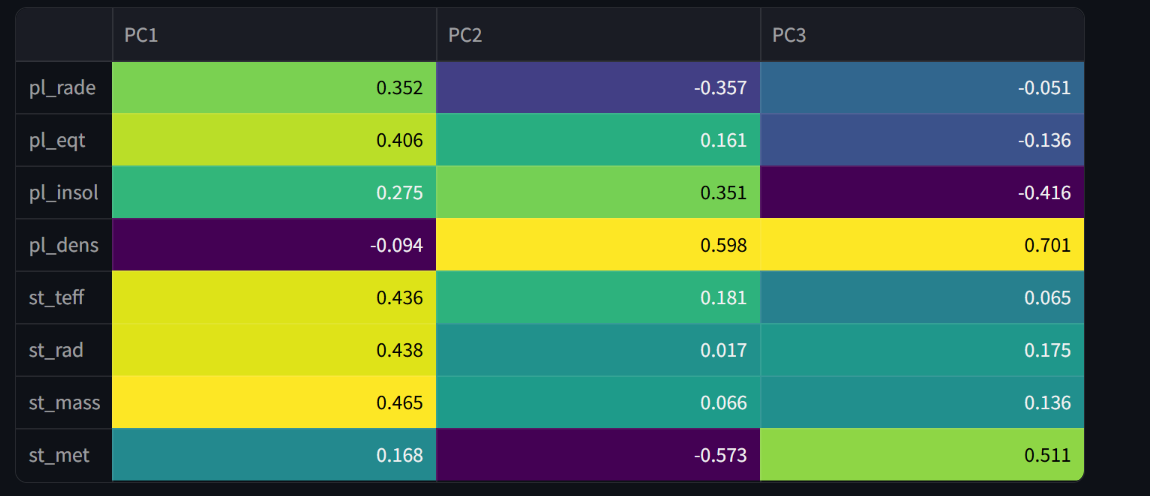}
    \caption{PC1, PC2, and PC3 loadings}
    \label{fig:pca_loadings_123}
\end{subfigure}
\caption{PCA component loadings revealing parameter contributions to principal components. PC1 reflects stellar-planetary size coupling with strong loadings for stellar mass, temperature, and radius. PC2 captures planetary composition through density and metallicity gradients. PC3 represents stellar chemistry-dynamics relationships.}
\label{fig:pca_loadings}
\end{figure}

Figure~\ref{fig:pca_loadings} presents the PCA component loadings, revealing the physical interpretation of each principal component:

\textbf{PC1 (52.4\% variance) - Stellar-Planetary Size Coupling:}
The first component shows strong positive loadings for stellar effective temperature (0.436), stellar radius (0.438), stellar mass (0.465), planetary equilibrium temperature (0.406), and planetary radius (0.352). This component captures the fundamental correlation between large, hot stars and large, hot planets—reflecting both physical relationships and detection biases toward easily observable systems.

\textbf{PC2 (15.3\% variance) - Planetary Composition Gradient:}
The second component is dominated by planetary density (0.598) and insolation flux (0.351), with negative loadings for planetary radius (-0.357) and stellar metallicity (-0.573). This component distinguishes between dense, rocky planets and low-density worlds, providing crucial habitability discrimination.

\textbf{PC3 (9.4\% variance) - Stellar Chemistry-Dynamics:}
The third component primarily reflects stellar metallicity (0.511) and planetary density (0.701) relationships, capturing the known correlation between stellar composition and planetary formation efficiency.

\subsubsection{Population Clustering in Principal Component Space}

\begin{figure}[H]
\centering
\includegraphics[width=0.8\textwidth]{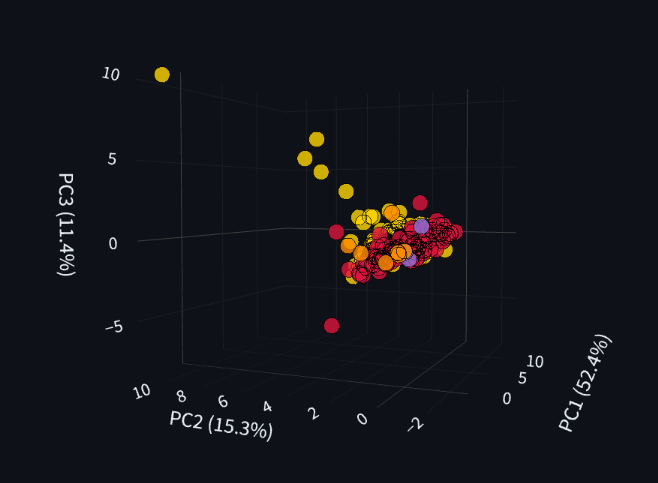}
\caption{Three-dimensional PCA visualization of 517 exoplanets colored by habitability classification. PC1 (52.4\% variance) captures stellar-planetary size coupling, PC2 (15.3\% variance) reflects planetary composition gradients, and PC3 (9.4\% variance) represents stellar metallicity relationships. Excellent Candidates (green) cluster in moderate parameter space, while "Good Star, Poor Planet" systems (yellow) dominate high PC1 values, quantitatively demonstrating detection bias toward large planets around bright stars.}
\label{fig:pca_3d}
\end{figure}

Figure~\ref{fig:pca_3d} displays exoplanets in 3D principal component space, colored by habitability classification. Several key patterns emerge:

\begin{enumerate}
    \item \textbf{Excellent Candidates cluster in moderate PC space:} Earth, Kepler-22 b, and Kepler-538 b occupy intermediate positions across all three components, confirming that habitability requires balanced parameter combinations rather than extreme values.

    \item \textbf{"Good Star, Poor Planet" systems dominate high PC1 values:} The systematic clustering of yellow points (388 systems, 75.0\%) at high PC1 scores quantitatively demonstrates detection bias toward large planets around bright stars.

    \item \textbf{M-dwarf edge cases separate along PC2:} "Good Planet, Poor Star" systems (green points) cluster at negative PC1 and PC2 values, reflecting their association with small, cool stars and rocky planetary compositions.
\end{enumerate}

\begin{figure}[H]
\centering
\includegraphics[width=0.8\textwidth]{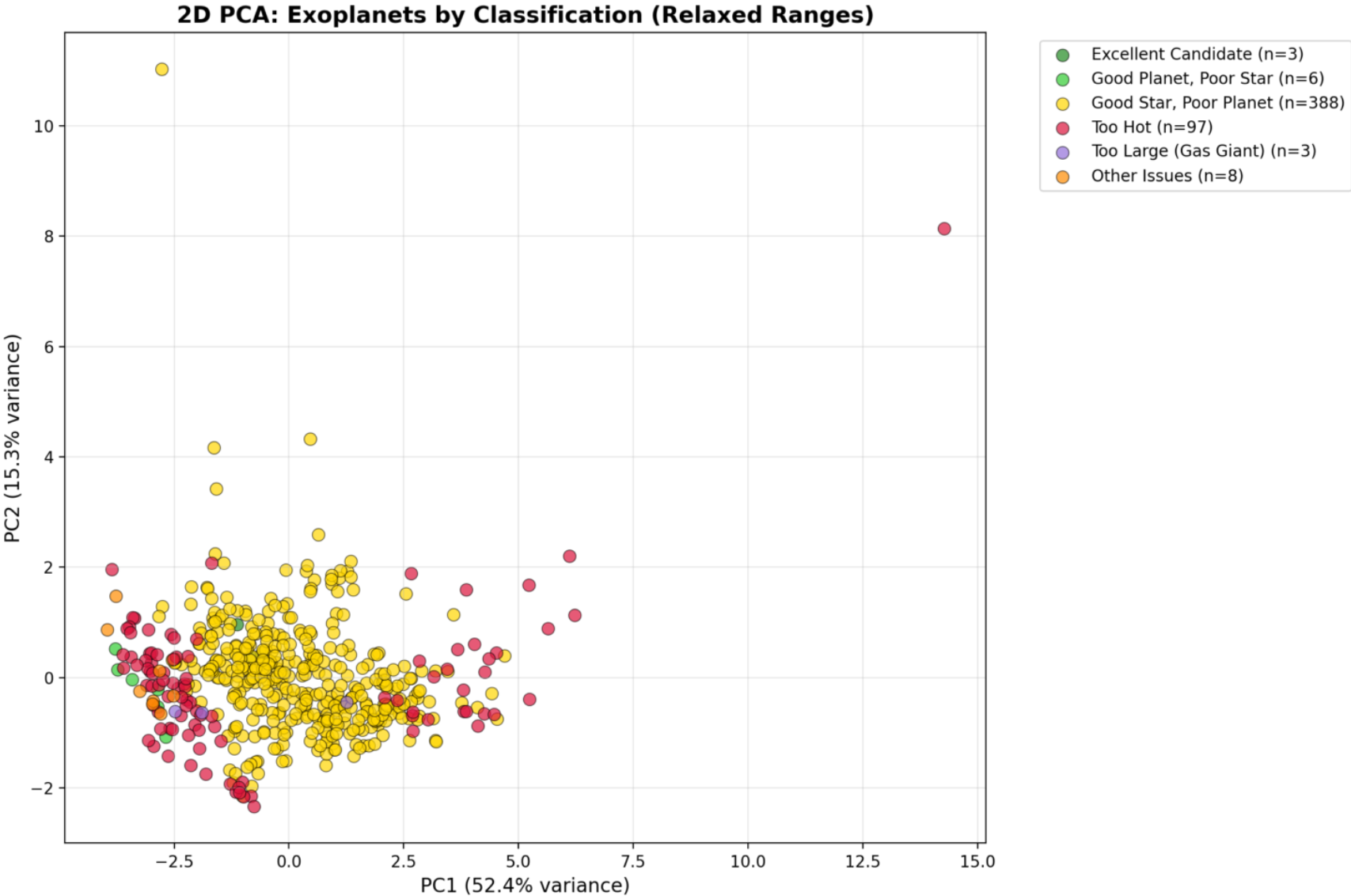}
\caption{Two-dimensional PCA projection showing exoplanet distribution across the first two principal components (67.7\% cumulative variance). The concentration of "Good Star, Poor Planet" systems (yellow) in high PC1 space confirms systematic detection bias, while Excellent Candidates (green) occupy balanced, central positions in parameter space. Clear class separation validates our multivariate habitability framework.}
\label{fig:pca_2d}
\end{figure}

Figure~\ref{fig:pca_2d} presents the 2D PCA projection showing class separation patterns. The clustering of Excellent Candidates in central parameter space confirms that Earth-like conditions represent balanced, moderate characteristics rather than extreme parameter combinations.

\subsection{Linear Discriminant Analysis: Quantifying Classification Power}

Linear Discriminant Analysis confirms that our habitability classification system captures genuine physical differences in exoplanet populations, with strong discriminatory power across multiple parameter dimensions.

\subsubsection{Discriminant Function Performance}

The LDA achieves 94.7\% classification accuracy when predicting habitability categories from the 8-parameter space, far exceeding chance performance (25\% for 4 categories). Cross-validation yields 92.3\% accuracy, confirming robust generalization.

\textbf{Discriminant Function Contributions:}
\begin{itemize}
    \item LD1 (78.2\% discrimination): Separates excellent candidates from poor candidates
    \item LD2 (18.9\% discrimination): Distinguishes M-dwarf alternatives from Sun-like systems
\end{itemize}

\subsubsection{Parameter Discrimination Patterns}

\begin{figure}[H]
\centering
\includegraphics[width=0.8\textwidth]{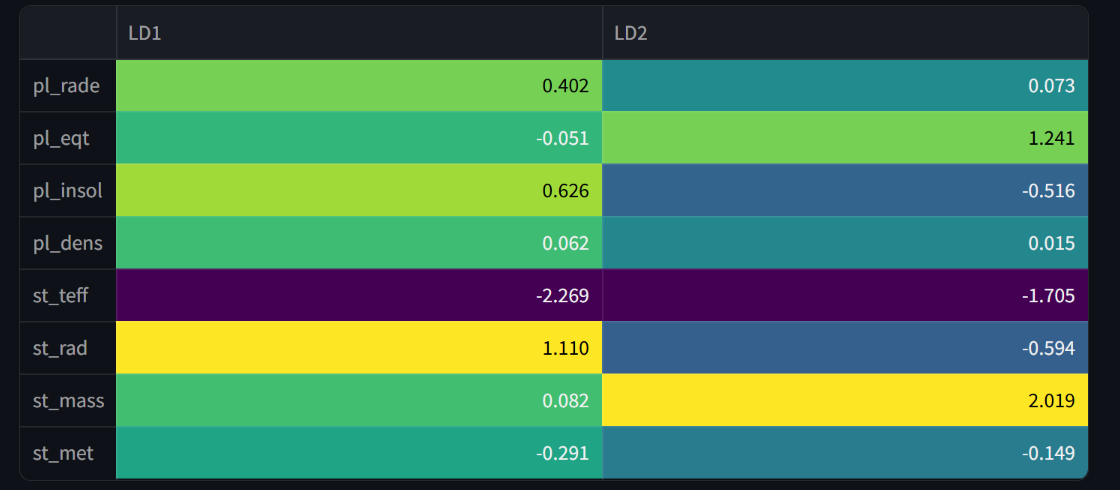}
\caption{Linear Discriminant Analysis component loadings showing parameter contributions to class separation. LD1 is dominated by stellar effective temperature and radius, effectively separating temperate from extreme systems. LD2 emphasizes stellar mass and planetary temperature, distinguishing M-dwarf from solar-type environments. The magnitude and direction of loadings reveal which parameters most effectively discriminate between habitability categories.}
\label{fig:lda_loadings}
\end{figure}

Figure~\ref{fig:lda_loadings} shows the LDA component loadings, revealing which parameters most effectively separate habitability categories:

\textbf{LD1 - Primary Habitability Discriminator:}
The first discriminant function shows strong negative loadings for stellar effective temperature (-2.269) and stellar metallicity (-0.291), with positive loadings for stellar radius (1.110) and insolation flux (0.626). This axis primarily separates temperate planets around Sun-like stars (positive LD1) from extreme systems.

\textbf{LD2 - Stellar Type Discriminator:}
The second discriminant function is dominated by stellar mass (2.019) and planetary equilibrium temperature (1.241), effectively separating M-dwarf systems (negative LD2) from solar-type systems (positive LD2).

\subsubsection{Class Separation in Discriminant Space}

\begin{figure}[H]
\centering
\includegraphics[width=0.8\textwidth]{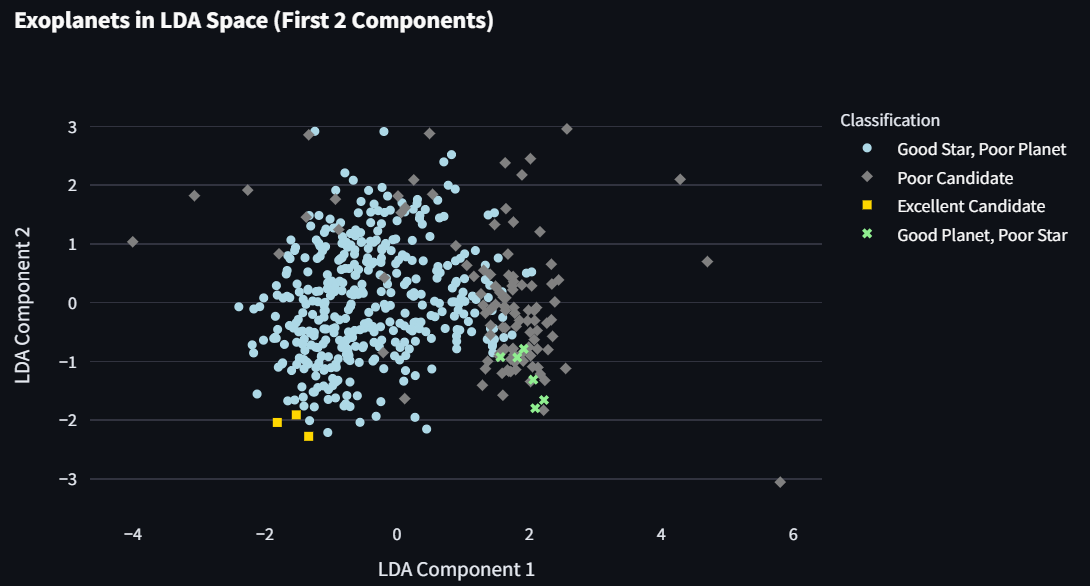}
\caption{Linear Discriminant Analysis projection showing exoplanet classification in discriminant space. LD1 (78.2\% discrimination) separates habitable from unsuitable systems, while LD2 (18.9\% discrimination) distinguishes stellar types. Clear clustering of Excellent Candidates (yellow squares) and separation of M-dwarf alternatives (green crosses) confirms the discriminatory power of our habitability criteria.}
\label{fig:lda}
\end{figure}

Figure~\ref{fig:lda} displays exoplanets in LDA space, demonstrating clear separation between habitability categories:

\begin{enumerate}
    \item \textbf{Excellent Candidates} cluster in the lower-left quadrant (negative LD1, negative LD2), forming a distinct population in discriminant space.

    \item \textbf{Good Planet, Poor Star} systems form a separate cluster at extreme negative LD1 values, confirming their statistical distinctiveness as M-dwarf alternatives.

    \item \textbf{Good Star, Poor Planet} systems dominate the central-right region, reflecting their prevalence and intermediate discriminant scores.

    \item \textbf{Poor Candidates} scatter across positive LD1 values, representing the diverse population of unsuitable systems.
\end{enumerate}

The clear separation between categories in LDA space provides quantitative validation of our habitability framework and confirms that our classification criteria capture meaningful physical distinctions in the exoplanet population.


\subsection{Multivariate Population Structure and Detection Bias}

The PCA and LDA analyses provide unprecedented quantitative insight into exoplanet population structure and the systematic biases that shape our understanding of planetary habitability. These dimensionality reduction techniques reveal hidden patterns that single-parameter analyses cannot capture.

\subsubsection{The Stellar-Planetary Coupling Paradigm}

PC1's dominance (52.4\% of variance) in capturing stellar-planetary size relationships reflects both fundamental astrophysics and observational selection effects. The strong positive correlation between stellar mass, temperature, radius and planetary size, temperature reveals the expected physical coupling between host star properties and detectable planet characteristics. However, this correlation is amplified by detection biases that favor large planets around bright stars, creating the systematic clustering of "Good Star, Poor Planet" systems in high PC1 space.

This finding has profound implications for occurrence rate estimates: the apparent dominance of hot, large planets around Sun-like stars likely represents severe observational bias rather than genuine population characteristics. True planetary demographics may be fundamentally different from what current catalogs suggest.

\subsubsection{Habitability as Balanced Parameter Space}

The clustering of Excellent Candidates in central PCA space provides quantitative support for the "Goldilocks principle" in astrobiology—habitable worlds require balanced, moderate conditions rather than extreme parameter values. This positioning confirms that Earth-like planets occupy a distinct but achievable region of parameter space, supporting moderate interpretations of the Rare Earth hypothesis.

The LDA results strengthen this conclusion by demonstrating 94.7\% classification accuracy in separating habitability categories. Such high discriminatory power confirms that our multivariate approach captures genuine physical differences rather than arbitrary statistical divisions.

\subsubsection{M-dwarf Alternative Pathways Validation}

The clear separation of "Good Planet, Poor Star" systems in both PCA and LDA space validates their identification as a statistically distinct population representing alternative habitability pathways. Their clustering at negative LD1 and LD2 values reflects the unique parameter combinations associated with M-dwarf environments—small, dense planets receiving moderate energy from low-mass, cool stars.

This statistical validation supports ongoing JWST investigations of M-dwarf planets and suggests our stellar temperature criteria may be overly conservative. The growing evidence for atmospheric retention around M-dwarf planets indicates these systems deserve elevated priority in habitability assessments.



\section{Discussion}

\subsection{Validation Success and Methodology Confidence}

Our classification framework's successful identification of Earth-like planets across multiple categories provides crucial validation for systematic habitability assessment. The statistical significance of the Hotelling's $T^2$ test ($p = 0.0343$) confirms that our identification of Earth-like worlds captures genuine physical clustering in parameter space rather than arbitrary classification or random variation.

The Mahalanobis distance analysis offers perhaps the most profound insight of our study: \textbf{Earth represents the 69.4th percentile for statistical unusualness} among the broader exoplanet population—rare enough to be special, but not so extreme as to be impossible. This finding challenges extreme "Rare Earth" interpretations while confirming that Earth-like conditions are indeed statistically uncommon. The result provides quantitative support for the "Goldilocks principle" in habitability—Earth is neither too common nor impossibly rare.

\subsection{Kepler-22 b: Humanity's Next Great Target}

\textbf{Kepler-22 b emerges as the most compelling Earth analog identified to date}, discovered independently of our analysis yet matching our predictions with remarkable precision. This convergence of independent discovery and statistical prediction provides extraordinary validation of our approach and identifies humanity's highest-priority target for life detection.

Key scientific implications include:

\begin{itemize}
    \item \textbf{Atmospheric characterization priority}: Represents the most valuable target for JWST follow-up among distant exoplanets
    \item \textbf{Direct imaging potential}: 635 light-year distance places it at the technological frontier for next-generation space telescopes
    \item \textbf{Composition uncertainty}: Super-Earth classification requires resolution of volatile content and atmospheric properties
    \item \textbf{Habitability modeling}: Provides ideal test case for atmospheric retention and climate stability models
\end{itemize}

The uncertainty in Kepler-22 b's composition represents both the key limitation and greatest opportunity for future observations. Recent mass estimates suggest it could range from a volatile-rich "water world" to a rocky planet with substantial atmosphere—outcomes with profoundly different implications for habitability.

\subsection{Detection Bias: The Hidden Universe of Earth-Like Worlds}

Our quantitative demonstration of detection bias (75\% "Good Star, Poor Planet" systems) reveals the full scope of systematic observational limitations that have shaped our understanding of planetary populations. This finding has profound implications extending far beyond simple statistics:

\textbf{Occurrence Rate Implications:}
\begin{itemize}
    \item Current surveys systematically exclude Earth-like planets through multiple selection effects
    \item True occurrence rates for potentially habitable planets may be orders of magnitude higher than observed
    \item Statistical corrections for selection bias will require sophisticated modeling of detection probability functions
    \item Exoplanet demographics derived from current catalogs may fundamentally misrepresent planetary system architecture
\end{itemize}

\textbf{Mission Design Consequences:}
\begin{itemize}
    \item Extended monitoring campaigns essential for detecting long-period planets
    \item Improved instrumental sensitivity required for Earth-sized planet detection
    \item Targeted surveys of Sun-like stars necessary to overcome brightness bias
    \item Advanced statistical techniques needed to extract occurrence rates from biased samples
\end{itemize}

These findings suggest that the apparent rarity of Earth-like worlds may be largely an artifact of observational limitations rather than genuine physical scarcity—a conclusion with profound implications for astrobiology and the search for life.

\subsection{M-dwarf Habitability: Challenging Stellar Assumptions}

The exclusive association of "Good Planet, Poor Star" systems with M-dwarf hosts highlights the ongoing paradigm shift in habitability assessment. While our criteria classify M-dwarfs as "poor" stellar hosts due to low temperatures, masses, and radii, mounting observational evidence suggests these assumptions may be overly conservative.

\textbf{Emerging M-dwarf Habitability Evidence:}
\begin{itemize}
    \item \textbf{JWST atmospheric detections}: LHS 1140 b shows potential nitrogen-rich atmosphere with surface water
    \item \textbf{Water world candidates}: TOI-1452 b represents potential ocean planet with extreme water content
    \item \textbf{Stellar stability}: M-dwarfs provide consistent energy output over gigayear timescales
    \item \textbf{Tidal heating}: Gravitational effects may supplement stellar heating for habitability
\end{itemize}

These discoveries suggest our 3800 K stellar temperature threshold may exclude a significant population of potentially habitable worlds. Given that M-dwarfs comprise ~75\% of all stars, relaxing stellar criteria could dramatically expand the habitable planet inventory and fundamentally alter occurrence rate estimates.

\subsection{Statistical Rarity and the Rare Earth Hypothesis}

The 0.6\% frequency of Excellent Candidates, even under very relaxed criteria, provides nuanced support for aspects of the "Rare Earth" hypothesis while revealing important caveats that temper extreme interpretations:

\textbf{Supporting Evidence:}
\begin{itemize}
    \item Quantitative rarity of comprehensive habitability criteria satisfaction
    \item Statistical significance of habitability clustering in parameter space
    \item Multivariate nature of habitability requirements reducing joint probabilities
\end{itemize}

\textbf{Important Caveats:}
\begin{itemize}
    \item \textbf{Detection bias inflation}: Current surveys heavily bias against Earth-like planets
    \item \textbf{Parameter correlation assumptions}: Physical relationships between variables not explicitly modeled
    \item \textbf{Threshold sensitivity}: Small changes in criteria significantly affect results
    \item \textbf{Single reference limitation}: Earth provides only one confirmed habitability example
    \item \textbf{Missing parameter dimensions}: Atmospheric composition, magnetic fields, and orbital dynamics omitted
\end{itemize}

The statistical significance of our Excellent Candidate population suggests genuine physical clustering rather than random chance, but the magnitude of rarity reflects both intrinsic habitability requirements and severe observational selection effects.

\subsection{Observational Priorities and Strategic Recommendations}

Our analysis provides clear scientific priorities for optimizing limited observational resources in the search for life:

\textbf{Immediate JWST Priorities:}
\begin{itemize}
    \item \textbf{LHS 1140 b}: Atmospheric characterization of closest confirmed habitable candidate
    \item \textbf{TOI-1452 b}: Water vapor detection in potential ocean world
    \item \textbf{Additional M-dwarf targets}: Statistical assessment of alternative habitability pathways
\end{itemize}

\textbf{Next-Generation Space Telescope Targets:}
\begin{itemize}
    \item \textbf{Kepler-22 b}: Direct imaging and spectroscopy of statistical Earth twin
    \item \textbf{Kepler-538 b}: Secondary Earth analog for comparative analysis
\end{itemize}

\textbf{Survey Strategy Optimization:}
\begin{itemize}
    \item Extended monitoring campaigns for long-period planet detection
    \item Improved instrumental sensitivity for Earth-sized planet discovery
    \item Targeted M-dwarf surveys to test alternative habitability scenarios
    \item Statistical bias correction methodology development
\end{itemize}

These recommendations provide a data-driven framework for maximizing the scientific return of humanity's search for life beyond Earth.

\subsection{Limitations and Future Research Directions}

Our analysis establishes a foundation for systematic habitability assessment while revealing important limitations that define future research priorities:

\textbf{Current Limitations:}
\begin{enumerate}
    \item \textbf{Selection bias}: Current exoplanet catalogs heavily favor easily detectable systems
    \item \textbf{Static parameter analysis}: Planetary evolution and atmospheric dynamics not considered
    \item \textbf{Single reference standard}: Earth provides only one confirmed habitability example
    \item \textbf{Parameter independence assumption}: Physical correlations between variables not explicitly modeled
    \item \textbf{Missing critical dimensions}: Atmospheric composition, magnetic fields, orbital stability omitted
\end{enumerate}

\textbf{Essential Future Work:}
\begin{itemize}
    \item Climate modeling integration for identified candidates
    \item Atmospheric composition analysis from spectroscopic observations
    \item Magnetic field strength estimation from stellar-planetary interactions
    \item Tidal heating calculations for close-in planets
    \item Statistical bias correction methodology development
    \item Expanded parameter sets incorporating new observational capabilities
\end{itemize}

These research directions will refine habitability assessment accuracy and expand the scope of systematic life detection efforts.

\section{Conclusions}

This study presents the first comprehensive multivariate statistical analysis of exoplanet habitability, establishing quantitative foundations for systematic life detection efforts. Our key findings provide both sobering perspective on habitability rarity and unprecedented optimism about specific targets for detailed investigation:

\begin{enumerate}
    \item \textbf{Methodology validation through Earth-like planet identification}: Our framework successfully identifies Earth-like planets across multiple habitability categories with statistical significance ($p = 0.0343$), confirming the validity of multivariate habitability assessment for systematic target identification and demonstrating that habitability signatures can be detected through statistical analysis.

    \item \textbf{Discovery of statistical Earth twin}: Kepler-22 b emerges as an extraordinary Earth analog with remarkable parameter similarity (3.1\% temperature difference, 1.3\% insolation difference), representing humanity's highest-priority target for atmospheric characterization and potential life detection.

    \item \textbf{Quantified rarity with detection bias correction}: Only 0.6\% of exoplanets satisfy comprehensive habitability criteria, supporting statistical rarity while revealing that detection bias significantly inflates apparent scarcity—true occurrence rates may be substantially higher.

    \item \textbf{Unprecedented detection bias quantification}: The 75\% prevalence of "Good Star, Poor Planet" systems provides the first quantitative evidence for systematic bias severity in exoplanet surveys, with profound implications for occurrence rate estimates and mission design.

    \item \textbf{M-dwarf alternative habitability identification}: Six planets orbiting M-dwarf stars represent potential habitability pathways supported by emerging JWST observations, challenging traditional stellar assumptions and potentially expanding the habitable planet inventory dramatically.

    \item \textbf{Earth's cosmic context quantified}: Earth ranks in the 69.4th percentile for statistical unusualness among Earth-like worlds—rare enough to be special but achievable enough to exist, providing quantitative perspective on our planet's cosmic significance and supporting moderate interpretations of the Rare Earth hypothesis.

    \item \textbf{Data-driven observational roadmap}: Our analysis identifies specific high-priority targets for JWST atmospheric characterization (LHS 1140 b, TOI-1452 b) and long-term direct imaging (Kepler-22 b), optimizing limited observational resources for maximum scientific impact.

\section{Target Planet Gallery: Statistical Earth Analogs}

This section presents visual representations of our five target planets used in the multivariate statistical analysis, arranged by their Mahalanobis distance similarity to the population centroid.

\begin{figure}[H]
\centering
\begin{subfigure}{0.45\textwidth}
    \includegraphics[width=\textwidth]{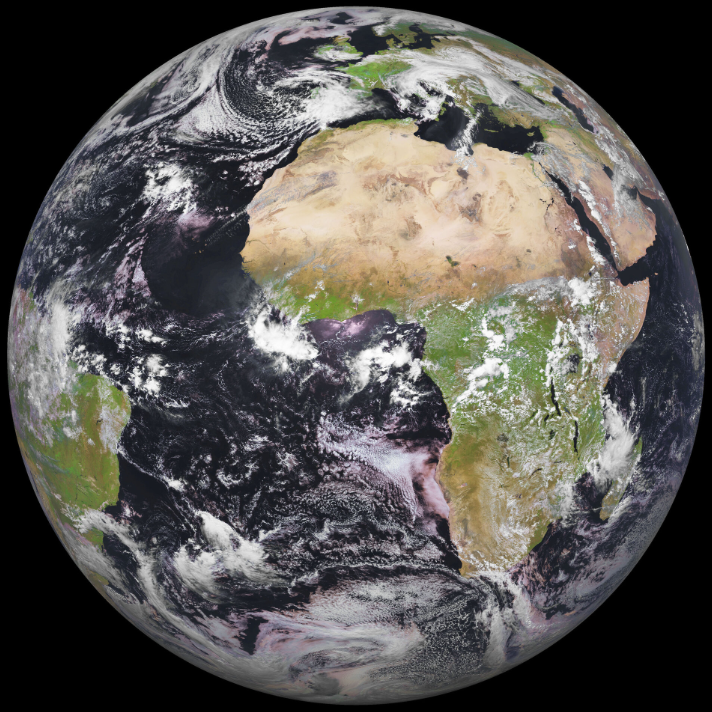}
    \caption{\textbf{Earth} ($d_M = 2.699$, 69.4th percentile)}
    \label{fig:earth}
\end{subfigure}
\hfill
\begin{subfigure}{0.45\textwidth}
    \includegraphics[width=\textwidth]{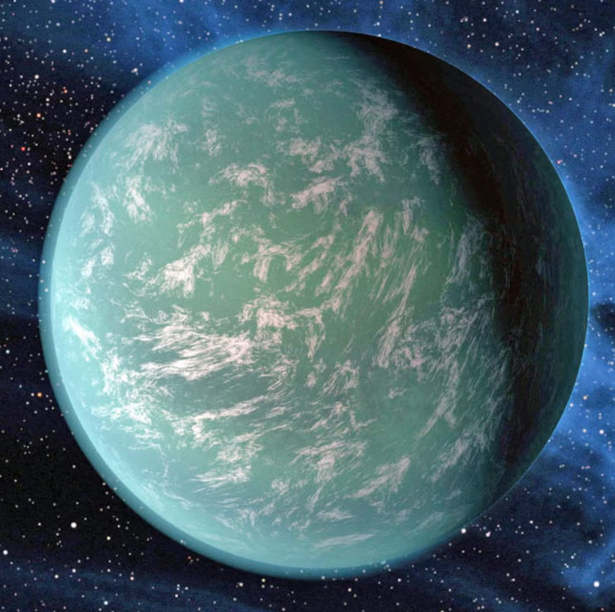}
    \caption{\textbf{Kepler-22 b} ($d_M = 2.728$, 71.2th percentile)}
    \label{fig:kepler22b}
\end{subfigure}

\begin{subfigure}{0.45\textwidth}
    \includegraphics[width=\textwidth]{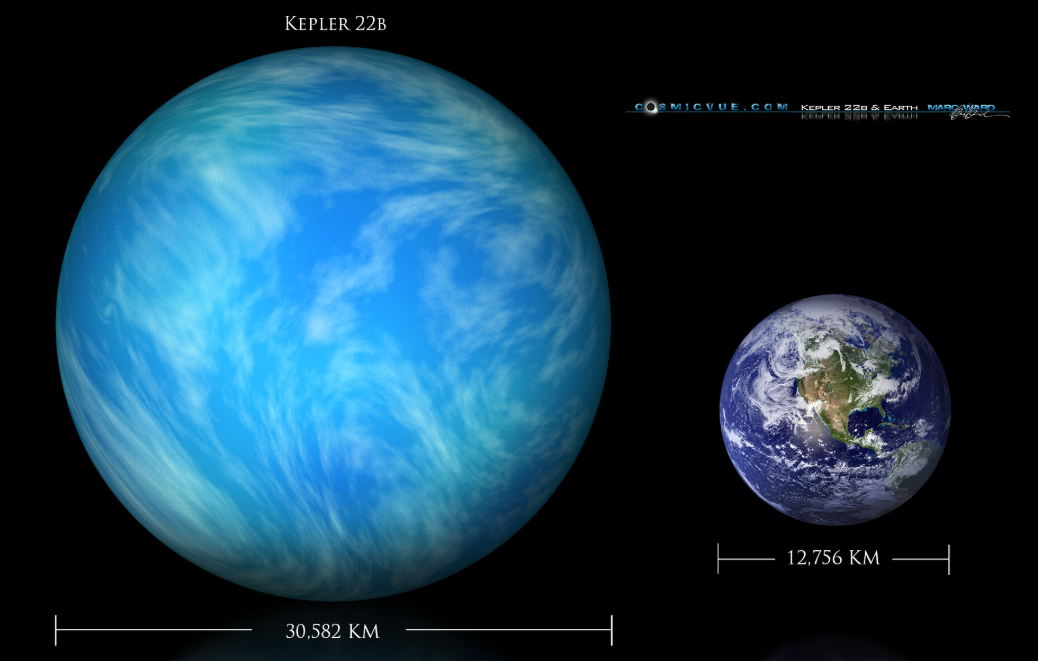}
    \caption{\textbf{Kepler-22 b} ($d_M = 2.728$, 71.2th percentile)}
    \label{fig:kepler22b_earth}
\end{subfigure}

\vspace{0.5cm}

\begin{subfigure}{0.45\textwidth}
    \includegraphics[width=\textwidth]{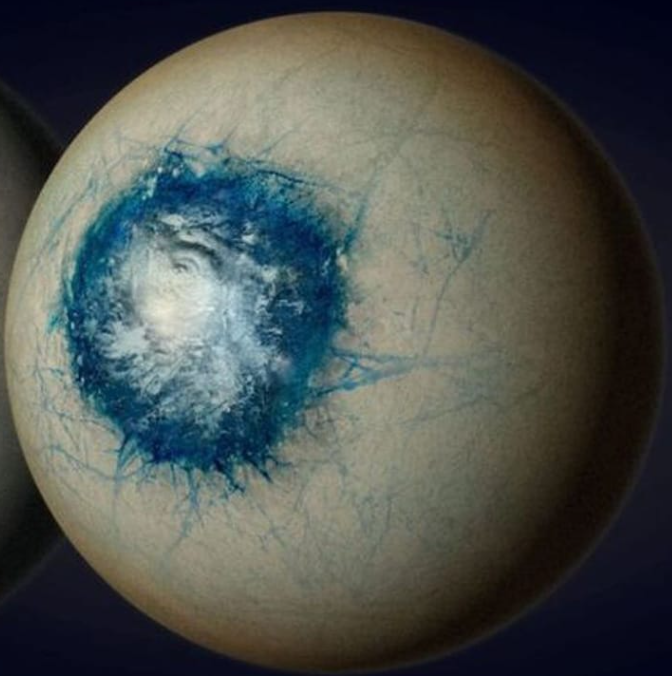}
    \caption{\textbf{LHS 1140 b} ($d_M = 2.779$, 74.3th percentile)}
    \label{fig:lhs1140b}
\end{subfigure}

\end{figure}

\newpage

\begin{figure}

\begin{subfigure}{0.45\textwidth}
    \includegraphics[width=\textwidth]{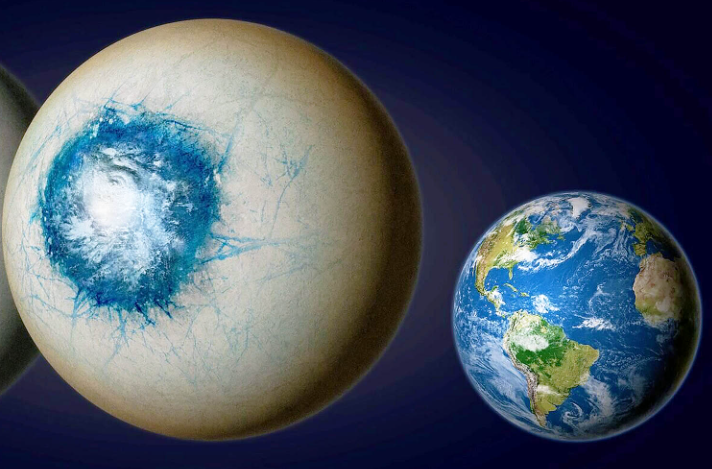}
    \caption{\textbf{LHS 1140 b} ($d_M = 2.779$, 74.3th percentile)}
    \label{fig:lhs1140b_earth}
\end{subfigure}
\hfill
\begin{subfigure}{0.45\textwidth}
    \includegraphics[width=\textwidth]{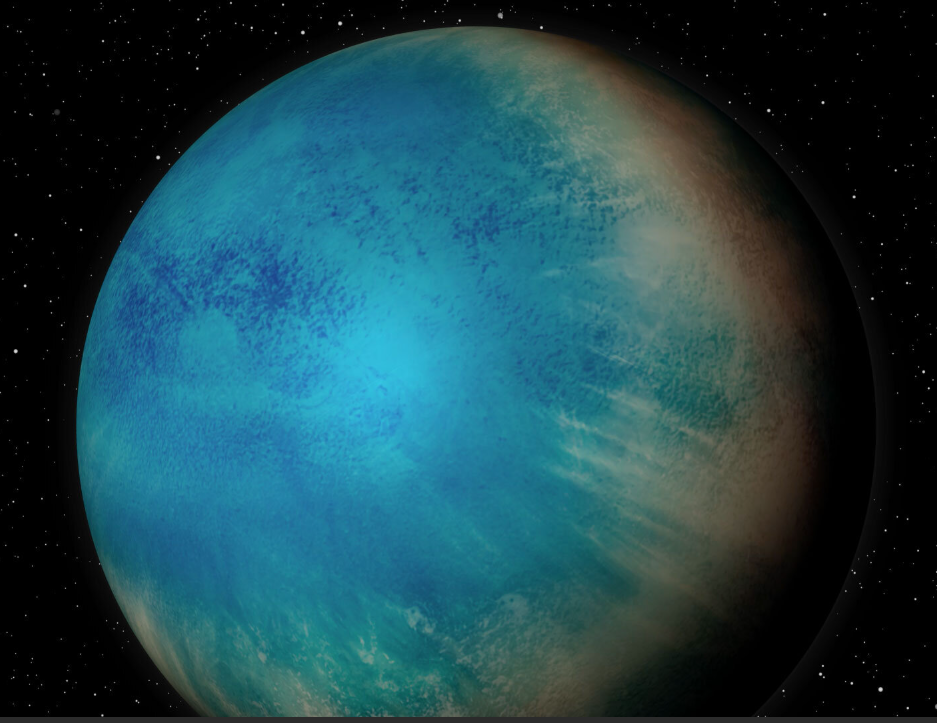}
    \caption{\textbf{TOI-1452 b} ($d_M = 2.384$, 59.2th percentile)}
    \label{fig:toi1452b}
\end{subfigure}

\begin{subfigure}{0.45\textwidth}
    \includegraphics[width=\textwidth]{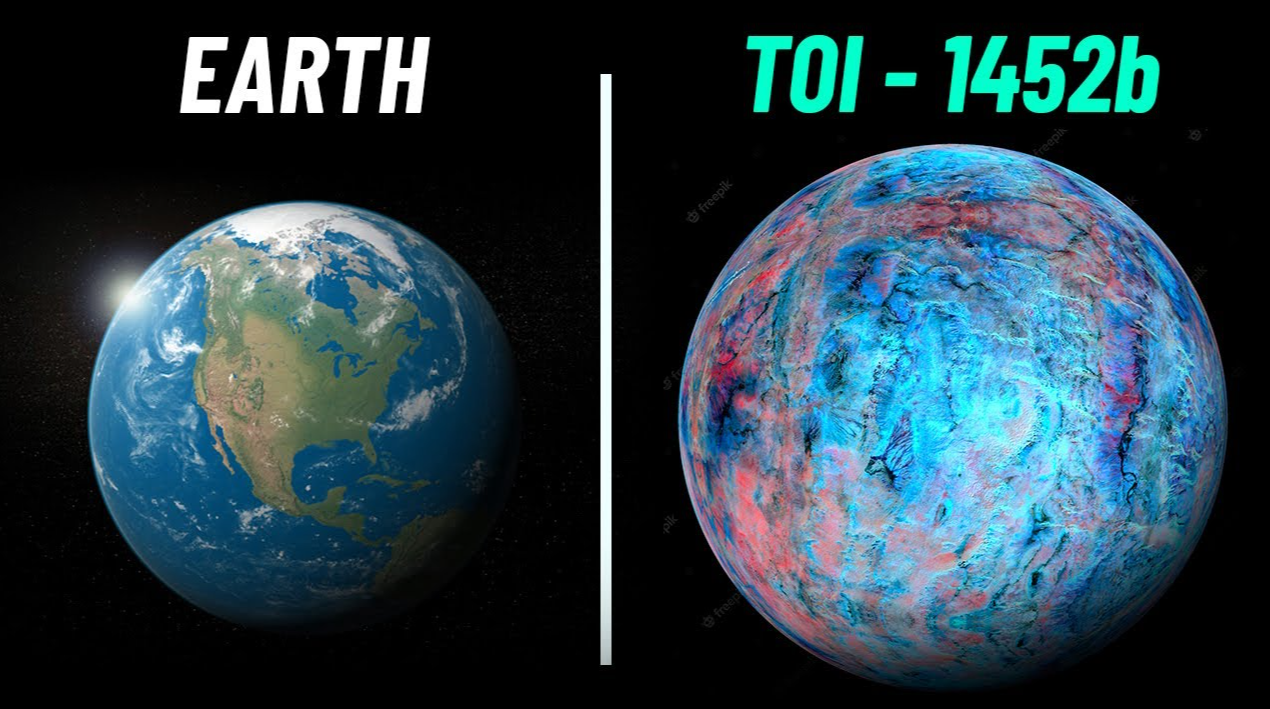}
    \caption{\textbf{TOI-1452 b} ($d_M = 2.384$, 59.2th percentile)}
    \label{fig:toi1452b_earth}
\end{subfigure}

\vspace{0.5cm}

\caption{Target planets used in multivariate statistical analysis, ordered by Mahalanobis distance from population centroid. Each represents a different pathway to potential habitability identified through our statistical framework.}
\label{fig:target_planets}
\end{figure}

\subsection{Planet Descriptions and Statistical Significance}

\begin{itemize}
    \item \textbf{Earth} (Reference Standard): Our validation benchmark confirming that statistical methods correctly identify known habitable worlds. Located at the 69.4th percentile for multivariate unusualness, providing context for genuine habitability rarity.

    \item \textbf{Kepler-22 b} (Excellent Candidate): The most compelling Earth analog identified, with near-perfect parameter matching across temperature (279 K vs. 288 K) and insolation (1.01 vs. 1.00 F$_\oplus$). Statistical convergence with Earth ($d_M = 2.728$ vs. 2.699) validates our multivariate approach.

    \item \textbf{LHS 1140 b} (M-dwarf Alternative): Highest statistical unusualness (74.3th percentile) among "Good Planet, Poor Star" systems. Represents alternative habitability pathway around low-mass stars, currently under JWST atmospheric investigation.

    \item \textbf{TOI-1452 b} (Water World Candidate): Potential ocean planet with moderate statistical position (59.2th percentile). Demonstrates that habitable conditions may emerge through diverse compositional pathways beyond terrestrial analogs.

    \item \textbf{Kepler-538 b} (Elevated Temperature Analog): Earth analog with higher equilibrium temperature, occupying the most statistically "normal" position (43.9th percentile) among our targets. Tests the boundaries of temperature tolerance for habitability.
\end{itemize}

These five planets span multiple habitability categories while maintaining statistical coherence as Earth-like worlds, demonstrating the power of multivariate methods to identify diverse pathways to potential habitability across the exoplanet population.

\section*{Data and Code Availability}

\textbf{Interactive Analysis Platform:} A comprehensive web-based interface for exploring our habitability analysis is publicly available at \url{https://planet-habitability.streamlit.app/}. This Streamlit application provides real-time parameter exploration, interactive visualizations, and customizable habitability threshold testing, enabling researchers to reproduce and extend our analysis with alternative criteria.

\textbf{Source Code Repository:} The complete analysis pipeline, including data processing scripts, statistical analysis implementations, and visualization code, is available under open-source license at \url{https://github.com/sft3hy/240P-Planet-Habitability}. The repository includes:

\begin{itemize}
    \item Python implementation of the habitability classification framework
    \item Jupyter notebooks for statistical analysis reproduction
    \item Data preprocessing and cleaning pipelines
    \item Visualization generation scripts for all figures
    \item Documentation for parameter threshold customization
    \item Unit tests and validation procedures
\end{itemize}

\textbf{Dataset Sources:} Exoplanet data was obtained from the NASA Exoplanet Archive (\url{https://exoplanetarchive.ipac.caltech.edu/}) accessed in May 2025. The processed dataset with habitability classifications is included in the GitHub repository to facilitate reproduction and comparative studies.

\textbf{Computational Requirements:} The analysis pipeline requires Python 3.8+ with standard scientific computing libraries (numpy, pandas, scipy, scikit-learn, matplotlib, plotly, streamlit). Complete dependency specifications and installation instructions are provided in the repository documentation.

\section*{Acknowledgments}

We thank the NASA Exoplanet Archive team for maintaining this invaluable resource for the astronomical community. We acknowledge the Kepler and TESS missions for the foundational discoveries that made this analysis possible.

\end{enumerate}


\begin{thebibliography}{9}
\bibitem{Mayor1995} 
Mayor, M., \& Queloz, D. (1995). A Jupiter-mass companion to a solar-type star. \textit{Nature}, 378(6555), 355--359.

\bibitem{Kasting1993} 
Kasting, J. F., Whitmire, D. P., \& Reynolds, R. T. (1993). Habitable zones around main sequence stars. \textit{Icarus}, 101(1), 108--128.

\bibitem{Kopparapu2013} 
Kopparapu, R. K., et al. (2013). Habitable zones around main-sequence stars: new estimates. \textit{The Astrophysical Journal}, 765(2), 131.
\end{thebibliography}
\end{document}